\documentclass[useAMS,usegraphicx,usenatbib]{mn2e}          

\usepackage{times}
\usepackage{flushend}

\newcommand{\Msun}{\ensuremath{\,{\rm M}_\odot}}                  
\newcommand{\Rsun}{\ensuremath{\,{\rm R}_\odot}}                  
\newcommand{\Mjup}{\ensuremath{\,{\rm M}_{\rm Jup}}}              
\newcommand{\Rjup}{\ensuremath{\,{\rm R}_{\rm Jup}}}              
\newcommand{\ms}{\,m\,s$^{-1}$}                                   
\newcommand{\chir}{\ensuremath{\chi_\nu^{\,2}}}                   
\newcommand{\mc}[1]{\multicolumn{2}{c}{#1}}

\newcommand{\reff}[1]{{#1}}
\newcommand{\refff}[1]{{#1}}

\title[Transit timing variations in the WASP-4 system]
      {Transit timing variations in the WASP-4 planetary system\thanks{Based on data collected by MiNDSTEp with the Danish 1.54\,m telescope at the ESO La Silla Observatory.}}

\author[Southworth et al.]
       {John Southworth\,$^{1}$, M. Dominik\,$^{2}$, U. G. J{\o}rgensen\,$^{3}$, M. I. Andersen\,$^{4}$, V. Bozza\,$^{5,6}$, \newauthor
        M. J. Burgdorf\,$^{7}$, G. D'Ago\,$^{8}$, S. Dib\,$^{9,10}$, R. {Figuera Jaimes}\,$^{11}$, Y. I. Fujii\,$^{3,4,12}$, S. Gill\,$^{1,13}$, \newauthor
        L. K. Haikala\,$^{14}$, T. C. Hinse\,$^{15}$, M. Hundertmark\,$^{10}$, E. Khalouei\,$^{16}$, H. Korhonen\,$^{17}$, \newauthor
        P. Longa-Pe{\~n}a\,$^{18}$, L. Mancini\,$^{19,10,20,21}$, N. Peixinho\,$^{22}$, M. Rabus\,$^{23,24}$, S. Rahvar\,$^{16}$, \newauthor
        S. Sajadian\,$^{25}$, J. Skottfelt\,$^{26}$, C. Snodgrass$^{27}$, P. Spyratos\,$^{1}$, J. Tregloan-Reed\,$^{18}$, \newauthor
        E. Unda-Sanzana\,$^{18}$, C. von Essen\,$^{28}$ \\
        $^{1}$\,Astrophysics Group, Keele University, Staffordshire, ST5 5BG, UK \\
        $^{2}$\,Centre for Exoplanet Science, SUPA, School of Physics \& Astronomy, University of St Andrews, North Haugh, St Andrews KY16 9SS, UK \\
        $^{3}$\,Niels Bohr Institute \& Centre for Star and Planet Formation, University of Copenhagen, {\O}ster Voldgade 5, 1350 Copenhagen, Denmark \\
        $^{4}$\,Niels Bohr Institute, Blegdamsvej 17, 2100 Copenhagen, Denmark \\
        $^{5}$\,Dipartimento di Fisica ``E.R. Caianiello'', Universit{\`a} di Salerno, Via Giovanni Paolo II 132, 84084, Fisciano, Italy \\
        $^{6}$\,Istituto Nazionale di Fisica Nucleare, Sezione di Napoli, Napoli, Italy \\
        $^{7}$\,Universit{\"a}t Hamburg, Department of Earth Sciences, Meteorological Institute, Bundesstra\ss{}e 55, 20146 Hamburg, Germany \\
        $^{8}$\,Instituto de Astrof{\'\i}sica, Facultad de F{\'\i}sica, Pontificia Universidad Cat\'olica de Chile, Av. Vicu\~na Mackenna 4860, 7820436 Macul, Santiago, Chile \\
        $^{9}$\,Laboratoire d'Astrophysique de Bordeaux, Universit\'{e} de Bordeaux, CNRS, B18N, all\'{e}e Geoffroy Saint-Hilaire, 33615, Pessac, France \\
        $^{10}$\,Max Planck Institute for Astronomy, K\"onigstuhl 17, D-69117 Heidelberg, Germany \\
        $^{11}$\,Facultad de Ingenier{\'\i}a y Tecnolog{\'\i}a Universidad San Sebast{\'\i}an, General Lagos 1163, Valdivia 5110693, Chile \\
        $^{12}$\,Institute for Advanced Research and Department of Physics, Nagoya University, Furo-cho, Chikusa-ku, Nagoya, 464-8601, Japan \\
        $^{13}$\,Department of Physics, University of Warwick, Gibbet Hill Road, Coventry CV4 7AL, UK \\
        $^{14}$\,Instituto de Astronom{\'\i}a y Ciencias Planetarias de Atacama, Universidad de Atacama, Copayapu 485,  Copiapo, Chile \\
        $^{15}$\,Chungnam National University, Department of Astronomy and Space Science, 34134 Daejeon, Republic of Korea \\
        $^{16}$\,Department of Physics, Sharif University of Technology, PO Box 11155-9161 Tehran, Iran \\
        $^{17}$\,DARK, Niels Bohr Institute, University of Copenhagen, Lyngbyvej 2, 2100 Copenhagen, Denmark \\
        $^{18}$\,Centro de Astronom{\'\i}a (CITEVA), Universidad de Antofagasta, Avda. U. de Antofagasta 02800, Antofagasta, Chile \\
        $^{19}$\,Department of Physics, University of Rome Tor Vergata, Via della Ricerca Scientifica 1, I-00133 Rome, Italy \\
        $^{20}$\,INAF -- Osservatorio Astrofisico di Torino, via Osservatorio 20, I-10025 Pino Torinese, Italy \\
        $^{21}$\,International Institute for Advanced Scientific Studies (IIASS), Via G. Pellegrino 19, I-84019 Vietri sul Mare (SA), Italy \\
        $^{22}$\,CITEUC -- Center for Earth and Space Science Research of the University of Coimbra, 3040-004 Coimbra, Portugal \\
        $^{23}$\,Las Cumbres Observatory Global Telescope, 6740 Cortona Dr., Suite 102, Goleta, CA 93111, USA \\
        $^{24}$\,Department of Physics, University of California, Santa Barbara, CA 93106-9530, USA \\
        $^{25}$\,Department of Physics, Isfahan University of Technology, Isfahan 84156-83111, Iran \\
        $^{26}$\,Centre for Electronic Imaging, Department of Physical Sciences, The Open University, Milton Keynes, MK7 6AA, UK \\
        $^{27}$\,Institute for Astronomy, University of Edinburgh, Royal Observatory, Edinburgh EH9 3HJ, UK \\
        $^{28}$\,Stellar Astrophysics Centre, Department of Physics and Astronomy, Aarhus University, Ny Munkegade 120, 8000 Aarhus C, Denmark
        }

\begin{document} \maketitle 


\begin{abstract}
Transits in the planetary system WASP-4 were recently found to occur 80\,s earlier than expected in observations from the TESS satellite. We present 22 new times of mid-transit that confirm the existence of transit timing variations, and are well fitted by a quadratic ephemeris with \refff{period decay $dP/dt = -9.2 \pm 1.1$\,ms\,yr$^{-1}$.} \reff{We rule out instrumental issues, stellar activity and the Applegate mechanism as possible causes. The light-time effect is also not favoured due to the non-detection of changes in the systemic velocity. Orbital decay and apsidal precession are plausible but unproven.} WASP-4\,b is only the third hot Jupiter known to show transit timing variations to high confidence. \reff{We discuss a variety of observations of this and other planetary systems that would be useful in improving our understanding of WASP-4 in particular and orbital decay in general.}
\end{abstract}

\begin{keywords}
planetary systems --- stars: fundamental parameters --- stars: activity --- stars: individual: WASP-4
\end{keywords}


\section{Introduction}                                                                                                              \label{sec:intro}

Thousands of transiting planets \reff{are currently known}. Under the assumption of Keplerian motion, the transits are expected to recur with a strict periodicity. However, there is a range of phenomena that might change this behaviour, including stellar activity \reff{\citep{Oshagh+13aa}}, gravitational interactions \reff{\citep{HolmanMurray05sci}}, the light-time effect in multiple systems \reff{\citep{Woltjer22ban}}, and orbital decay due to tides \reff{\citep{Birkby+14mn}}. Because of this, extensive efforts have been devoted to the detection of transit timing variations (TTVs) \citep[e.g.][]{Gibson+10mn,Harpsoe+13aa,Maciejewski+18aca}. Many examples have been found for small planets observed using the \textit{Kepler} satellite \citep{Mazeh+13apjs} but only \emph{two} detections have been made for a planet outside the \textit{Kepler} field. \citet{Maciejewski+16aa} used transit timing measurements taken over 3.3\,yr to find a quadratic term in the orbital ephemeris of WASP-12 to a significance of 5$\sigma$ \citep[see also][]{Patra+17aj,Maciejewski+18aca}, and WASP-47\,b shows sinusoidal TTVs due to the presence of two smaller short-period planets \citep{Becker+15apj}.

In this work we show that WASP-4\,b \citep{Wilson+08apj} is the third giant planet known to exhibit detectable TTVs, with a significance of \refff{8.4}$\sigma$. This confirms preliminary suggestions from \citet[][hereafter B19]{Bouma+19aj} based on data from the TESS satellite \reff{\citep{Ricker+15jatis}}. \citet{Baluev+15mn} have also claimed a tentative detection of TTVs in WASP-4, in the form of a sinusoidal variation with a period near 5\,d; we do not confirm this detection. When needed, we will adopt the physical properties of the WASP-4 system given by \citet{Me12mn}. In \reff{brief}, the planet has mass 1.25\Mjup\ and radius 1.36\Rjup, and is on a 1.338\,d orbit around a star of mass 0.93\Msun\ and radius 0.91\Rsun. The star's effective temperature is 5500\,K \citep{Mortier+13aa} and it shows magnetic activity in the form of dark starspots \citep{Sanchis+11apj}. No companion stars have been found in high-resolution imaging studies \citep{Ngo+15apj,Evans+16aa} and no long-term trend in systemic velocity has been identified \citep{Knutson+14apj}.


\section{Observations}

\begin{figure*} \includegraphics[width=\textwidth,angle=0]{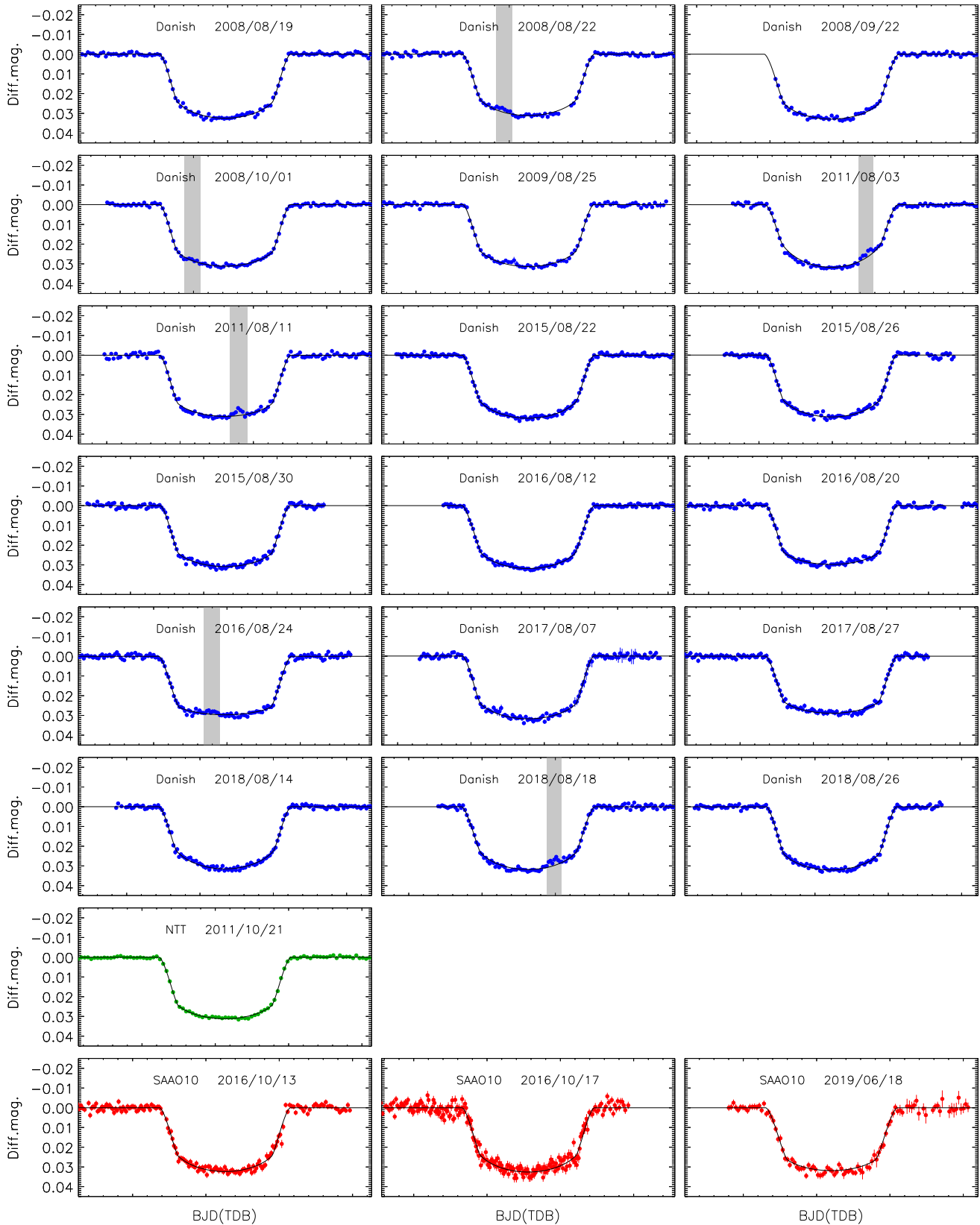}
\caption{\label{fig:lc} All new photometric data presented in this work. In each case
the x-axis shows 0.2\,d centred on the time of minimum measured for that transit. The
telescope and date are labelled on each panel. Starspot crossing events are visible
during totality \reff{in six of the light curves, and are indicated by the grey boxes.}
The {\sc jktebop} best fits are shown using black lines.} \end{figure*}

We have been monitoring transits \reff{in the WASP-4 system} for approximately a decade using the Danish 1.54\,m telescope and DFOSC imager at ESO La Silla, as a side-project of the MiNDSTEp Consortium \citep{Dominik+10an}. Four transits observed in the 2008 season were presented in \citet{Me+09mn2}, and we also observed one transit in 2009, two in 2011, three in 2015, three in 2016, two in 2017 and three in 2018. All \reff{observations} were obtained through a Cousins $R$ filter with the \reff{CCD unbinned and the} telescope significantly defocussed to enhance the photometric precision \citep{Me+09mn}. The scatters of the datasets versus transit fits (see below) range from 0.54 to 1.03 mmag with cadences of typically 140\,s.

We also observed one transit \reff{in the WASP-4 system} using the NTT and EFOSC2 imager \citep{Buzzoni+84msngr} with an SDSS $r$ filter (ESO filter \#784) \reff{on the night of 2011/10/21}. The telescope was defocussed, \reff{the CCD was not binned,} and only one comparison star was used due to the small field of view. The light curve is of very high quality \citep[see][]{TregloanMe13mn}, with a scatter of 0.47\,mmag.

\reff{Finally, we} observed three transits using the SAAO 1.0\,m telescope at Sutherland, South Africa: \reff{two in 2016 and one in 2019}. The STE4 CCD was used with a Cousins $R$ filter, 2$\times$2 binning to decrease the readout time, and the telescope defocussed. The resulting light curves have scatters of 1.5, 2.3 and 1.6 mmag. In the most recent run (2019 June) we cross-checked timings with another telescope at the observatory to confirm their reliability.

\subsection{Data reduction}

\reff{The raw data were reduced into light curves} using the {\sc defot} pipeline \citep{Me+09mn,Me+14mn} which implements aperture photometry, image motion tracking by cross-correlation with a reference image, and calculation of a differential-magnitude light curve. \reff{The light curve was obtained by simultaneously fitting the coefficients of a low-order polynomial versus time, and the weights of a set of comparison stars, to the data obtained outside transit.} In all cases we used a linear or quadratic polynomial and the coefficients were included as fitted parameters in all subsequent analyses.

\reff{The best light curve was obtained by manually iterating the aperture sizes and which comparison stars were included -- we found that the choices made here do affect the scatter of the light curve but have no significant effect on its shape.} We did not perform bias or flat-field calibrations because they have a negligible effect on DFOSC data \reff{except for an increase in the noise level of the light curve \citep{Me+14mn}.} All new data are shown in Fig.\,\ref{fig:lc}.

The timestamps were converted to the BJD(TDB) timescale using routines from \citet{Eastman++10pasp}. Manual time checks were performed for images in most observing sequences, and no discrepancies were noticed\footnote{The 2009 season with the Danish Telescope suffered from a timing problem \citep{Me+09apj,Me+10apj}. We have investigated this further and found that it did not affect any observations of WASP-4 presented here.}. We have previously found a good agreement between the timestamps on the Danish and MPG 2.2m telescopes \citep{Me+15mn}, supporting \reff{the reliability of both}.

\citet{Me+09mn2} used an early version of {\sc defot} so the 2008 data were re-reduced with the current version. This allowed the use of image motion tracking, additional comparison stars, and timestamps on the BJD(TDB) timescale. The re-reduced data supersede those from \citet{Me+09mn2}.


\section{Measurements of the transit times}

\begin{table} \centering \caption{\label{tab:tmin} Times of mid-transit for WASP-4 determined
in this work. The final column shows `y' if spot crossings are clearly visible in the data.}
\setlength{\tabcolsep}{4pt}
\begin{tabular}{lcrr@{\,$\pm$\,}lc} \hline
Telescope & \reff{Date} & \reff{Epoch} & \mc{BJD(TDB)} & Spot \\             
\hline
Danish    & \reff{2008/08/19} & \reff{$-$1351} & 2454697.79821 & 0.00010 &   \\      %
Danish    & \reff{2008/08/22} & \reff{$-$1348} & 2454701.81293 & 0.00011 & y \\      %
Danish    & \reff{2008/09/22} & \reff{$-$1325} & 2454732.59192 & 0.00017 &   \\      %
Danish    & \reff{2008/10/01} & \reff{$-$1319} & 2454740.62150 & 0.00007 & y \\      %
Danish    & \reff{2009/08/25} & \reff{$-$1073} & 2455069.82652 & 0.00009 &   \\      %
Danish    & \reff{2011/08/03} & \reff{ $-$544} & 2455777.75045 & 0.00014 & y \\      
Danish    & \reff{2011/08/11} & \reff{ $-$538} & 2455785.78063 & 0.00011 & y \\      
NTT       & \reff{2011/10/21} & \reff{ $-$485} & 2455856.70662 & 0.00007 &   \\      
Danish    & \reff{2015/08/22} & \reff{    562} & 2457257.83498 & 0.00007 &   \\      
Danish    & \reff{2015/08/26} & \reff{    565} & 2457261.84940 & 0.00010 &   \\      
Danish    & \reff{2015/08/30} & \reff{    568} & 2457265.86449 & 0.00010 &   \\      
Danish    & \reff{2016/08/12} & \reff{    828} & 2457613.80465 & 0.00010 &   \\      
Danish    & \reff{2016/08/20} & \reff{    834} & 2457621.83379 & 0.00011 &   \\      
Danish    & \reff{2016/08/24} & \reff{    837} & 2457625.84865 & 0.00010 & y \\      
SAAO 1.0m & \reff{2016/10/13} & \reff{    874} & 2457675.36302 & 0.00016 &   \\      
SAAO 1.0m & \reff{2016/10/17} & \reff{    877} & 2457679.37808 & 0.00018 &   \\      
Danish    & \reff{2017/08/07} & \reff{   1097} & 2457973.78903 & 0.00018 &   \\      
Danish    & \reff{2017/08/27} & \reff{   1112} & 2457993.86231 & 0.00014 &   \\      
Danish    & \reff{2018/08/14} & \reff{   1375} & 2458345.81705 & 0.00012 &   \\      
Danish    & \reff{2018/08/18} & \reff{   1378} & 2458349.83148 & 0.00010 & y \\      
Danish    & \reff{2018/08/26} & \reff{   1384} & 2458357.86123 & 0.00008 &   \\      
SAAO 1.0m & \reff{2019/06/18} & \reff{   1605} & 2458653.61010 & 0.00020 &   \\      
\hline \end{tabular} \end{table}

Each transit was fitted in isolation using the {\sc jktebop}\footnote{{\sc jktebop} is written in {\sc fortran77} and the source code is available at {\tt http://www.astro.keele.ac.uk/jkt/codes/jktebop.html}} code \citep[][and references therein]{Me13aa}. The fitted parameters were the transit midpoint $T_0$, the fractional radii of the two components ($r_{\rm A}=\frac{R_{\rm A}}{a}$ and $r_{\rm b}=\frac{R_{\rm b}}{a}$ where $R_{\rm A}$ is the radius of the star, $R_{\rm b}$ is the radius of the planet, and $a$ is the semimajor axis of the relative orbit) expressed as their sum and ratio, the orbital inclination, and the coefficients of the polynomial versus time. We modelled limb darkening using the quadratic law, fitted for the linear coefficient, and fixed the quadratic coefficient at an appropriate value.

Uncertainties on the transit times were obtained using Monte Carlo simulations and multiplied by $\sqrt{\chir}$ where $\chi^2_\nu$ is the reduced $\chi^2$ of the fit. This final step is necessary because the aperture photometry routine used in our pipeline tends to underestimate the observational uncertainties.
Table\,\ref{tab:tmin} gives the 22 transit times calculated in this work. To this dataset we added all timings from table~2 of B19, which includes 41 measurements from the literature and 18 timings from the TESS light curve.


\section{Transit timing analysis}

\begin{figure*}
\includegraphics[width=\textwidth,angle=0]{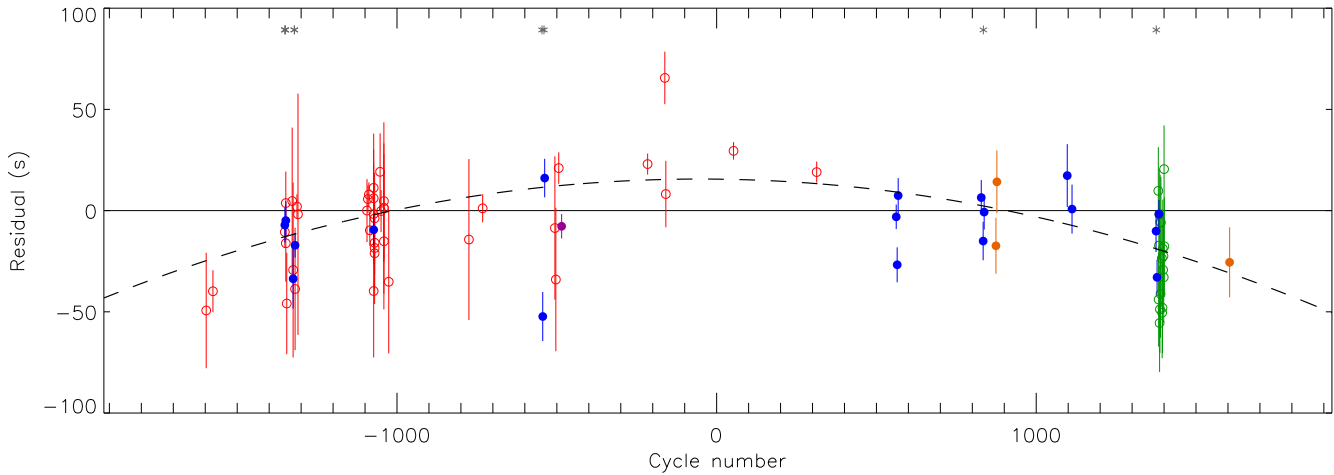}
\caption{\label{fig:minima} Plot of the residuals of the timings of mid-transit versus a
linear ephemeris. The results from this work are shown with filled circles: blue for the
Danish telescope, orange for the SAAO 1.0m, and purple for the NTT. Published results
are shown using open circles: red for the literature data collected by B19 and green for
the TESS timings from B19. The dotted line shows the difference between the best-fitting
linear and quadratic ephemerides. Grey asterisks have been placed near the top of the
figure to indicate transit times measured from a light curve with a spot crossing event.}
\end{figure*}

Our analysis was performed on the transit timings assembled above. We fitted a straight line to the times as a function of orbital cycle, using a zeropoint for cycle number near to the midpoint of the available data to avoid correlations between the parameters of the ephemeris (Fig.\,\ref{fig:minima}). This yielded a poor fit so we tried a quadratic ephemeris instead, finding a significant improvement. The best-fitting linear ephemeris is:
$$ T_0 = {\rm BJD(TDB)} \,\, 2\,456\,505.748953 (20) \, + \, 1.338231429 (20) E $$
with $\chir = 1.80$ \reff{and an rms scatter of 24.6\,s}. The bracketed quantities indicate the uncertainties in the preceding digits and have been increased by $\sqrt{\chir}$ to account for the imperfect fit.
The best-fitting quadratic ephemeris is:
\begin{eqnarray*}
T_0 & = & {\rm BJD(TDB)} \,\, 2\,456\,505.749133 (27)\\
    &   & \, + \, 1.338231408 (18) E \, - \, (1.95\!\pm\!0.23)\times10^{-10} E^2
\end{eqnarray*}
with $\chir = 1.41$ \reff{and an rms scatter of 19.9\,s. As above, the uncertainties have been increased by $\sqrt{\chir}$ to account for the imperfect fit.} We also fitted a cubic ephemeris but found it to be a negligible improvement on the quadratic ephemeris.


\refff{This quadratic coefficient corresponds to a period change of $\frac{dP}{dE} ~=~ (-3.91 \pm 0.46) \times 10^{-10}$ day per orbital cycle, and to a period derivative of
$$ \frac{dP}{dt} ~=~ -9.2 \pm 1.1 \mbox{~~ms yr}^{-1} $$
This is roughly $3\sigma$ lower than the value of $-12.6 \pm 1.2$\,ms\,yr$^{-1}$ found by B19. We investigated this by applying our analysis to only the transit times used by B19, which precisely reproduced their results. Therefore the difference in the period derivatives found by ourselves and by B19 is fully explained by the inclusion of additional transit times in the current study.}

The quadratic ephemeris has a much lower \chir\ than the linear ephemeris. To explore this further we calculated the Bayesian Information Criterion \citep{Schwarz78} to be 172.2 for the quadratic and 269.6 for the linear ephemeris, and the Akaike Information Criterion \citep{Akaike81} to be 165.1 for the quadratic and 264.8 for the linear ephemeris. All three statistical quantities give a significant improvement for the quadratic versus the linear ephemeris, meaning that we have detected TTVs in the WASP-4 system. The quadratic term is negative and measured to \refff{8.4}$\sigma$ significance.

The \chir\ of our preferred quadratic ephemeris is significantly above unity. We have found this situation to occur for most transit timing studies both in our own experience \citep[e.g.][]{Me+16mn} and in published studies \citep[e.g.][]{Baluev+15mn,Maciejewski+18ibvs}. The \refff{8.4}$\sigma$ significance level of the quadratic term already includes an allowance for this large \chir. Fig.\,\ref{fig:minima} shows that our new timings are consistent with published values in all cases where such a comparison can be drawn.

WASP-4\,A shows starspots and these are expected to add scatter to timing measurements both on theoretical \citep{Ballerini+12aa,Oshagh+13aa} and empirical \citep{Maciejewski+18aca} grounds. The obvious outlier at cycle $-544$ and residual $-64$\,s is affected by starspots. To investigate this we rejected the six timings clearly affected by starspot crossing events (Table\,\ref{tab:tmin}) and recalculated the ephemerides. We find that the \chir\ drops to 1.30 for the quadratic and 1.70 for the linear ephemeris, the quadratic term is not significantly changed, and its detection significance becomes 8.2$\sigma$. We conclude that spot activity increases the scatter of the transit timings but that TTVs are still detected to high confidence.

Visual inspection of Fig.\,\ref{fig:minima} suggested the possibility of a sinusoidal variation superimposed on the quadratic term. We therefore calculated a periodogram of the residuals of the best fit to the quadratic ephemeris, using the {\sc period04} package \citep{LenzBreger04iaus}. The periodogram contained no significant peaks, with the highest having a signal-to-noise ratio of 2.7. We find no significant peak near the period of 5.1\,d tentatively identified by \citet{Baluev+15mn}, so do not confirm this detection of sinusoidal variation. Further observations are needed to refine these conclusions.


\section{Possible causes of the transit timing variation}

In the previous section we have demonstrated that WASP-4 exhibits a quadratic variation in its transit times with a significance of \refff{8.4}$\sigma$. We now discuss possible causes of this effect.

\subsection{Starspots}

Transit times measured from light curves exhibiting spot crossing events can be biased away from the true time of midpoint because the fitted model (transit of a planet across a limb-darkened but otherwise uniform-brightness star) is no longer adequate \citep{Ballerini+12aa,Oshagh+13aa}. The bias depends on where the spot crossing event is, with those near the start or end of a transit having the strongest effect \reff{\citep{Oshagh+13aa}}. As activity levels on stars undergo cycles lasting years to decades \citep[e.g.][]{Baliunas+95apj}, these effects could conceivably mimic a slow TTV. We have \reff{visually inspected all transits presented in the current work in order to find those with clear spot crossing events. We have previously found visual methods to perform well in the identification of spot anomalies in transit light curves \citep{Mocnik++17mn}. We have} indicated the transits with spot crossing events on Fig.\,\ref{fig:minima}, and find no clear correlation with the residuals versus the quadratic ephemeris fit. The quadratic term in the ephemeris is also detected to almost the same significance if these timings are rejected from the ephemeris fit.

\refff{From Fig.\,\ref{fig:lc} we see that starspots are preferentially detected in WASP-4 towards the start or end of totality in the transit light curve. This is because the starspots move with the stellar surface at an approximately constant angular speed, and thus at a variable linear speed when projected into the plane of the sky. The starspots move faster when they are near the projected centre of the star, so are less likely to be found there. Conversely, smaller starspots may be preferentially found away from the limb of the star because limb darkening will attenuate their photometric signal and make their detection less likely. Both effects should be accounted for in detailed statistical studies of these phenomena.}

\citet{WatsonDhillon04mn} investigated the effect of starspots on timing measurements due to the Wilson depression\reff{: the phenomenon that starspots are slightly depressed into the surface of the star. They found that this effect was sufficient to cause a change in the measured time of an eclipse at the level of a few seconds for a white dwarf plus M-dwarf binary. The effect will be less than a second for WASP-4 due to the larger radius ratio, so is not capable of causing the TTV we have found.}

\subsection{Applegate mechanism}

Changes in the quadrupole moment of the star \citep{Applegate92apj} have been proposed as a possible cause of TTVs in planetary systems. \citet{WatsonMarsh10mn} calculated the amplitude of the TTVs for WASP-4 for three scenarios concerning the length of the activity cycle in the host star. The largest TTV amplitude was found to be 15.3\,s for a 50\,yr activity cycle, so is a factor of six too small to explain the observed TTVs.

\subsection{Orbital decay}

Tidal effects will cause the majority of known hot Jupiters to merge with their host stars on a timescale of several Gyr \citep{Levrard++09apj,Jackson++09apj}. This orbital decay will initially appear as a continuous decrease in their orbital period, causing a quadratic variation in the transit times. We followed the approach of \citet{Patra+17aj} to calculate the modified stellar tidal quality factor \reff{for the WASP-4 system}, finding $Q_\star^\prime = 10^{4.58 \pm 0.03}$.

\reff{A wide range of possible $Q_\star^\prime$ values have been found in previous studies, both theoretical and observational. The canonical value of $Q_\star^\prime$ is $10^6$ \citep[e.g.][]{OgilvieLin07apj}, a value of $10^{5.5}$ was found to be a good match to a sample of extrasolar planets \citep{Jackson++08apj2}, and a theoretical study by \citet{EssickWeinberg16apj} yielded $Q_\star^\prime \approx 10^5$ to $10^6$ for short-period hot Jupiters. But other studies disagree: theoretical work by \citet{PenevSasselov11apj} constrained $Q_\star^\prime$ to be $10^8$--$10^{9.5}$, an observational study by \citet{Penev+12apj} found $Q_\star^\prime > 10^7$, and \citet{CameronJardine18mn} used the distribution of orbital separations of the known population of transiting hot Jupiters to deduce $Q_\star^\prime = 10^{8.26 \pm 0.14}$.}

\reff{The value of $Q_\star^\prime$ is also expected to depend on the nature of the tidal perturbation and the internal structure of the star \citep{Ogilvie14araa,Penev+18aj}, on orbital period \citep{Barker11mn}, and on planet mass \citep{BarkerOgilvie10mn}. \citet{Barker11mn} found a dependence on orbital period of $Q_\star^\prime \propto P^{14/5}$, giving a value of approximately $10^{5.3}$ for the WASP-4 system.}

\reff{A tidal quality factor small enough to cause the TTV observed in WASP-4 is smaller than any determined in the observational and theoretical studies listed above. However, it is sufficiently close to the lower envelope of $Q_\star^\prime$ values, and this envelope is sufficiently large, that it is not currently possible to rule out orbital decay as a cause of the TTV. Further work is needed to better understand the possible values of $Q_\star^\prime$ and its dependence on system parameters such as planet mass and orbital period.}

\subsection{Light-time effect}

The presence of a third component in the system on a wider orbit than WASP-4\,b will cause a change in the measured transit timings due to the varying distance of \reff{the transiting system} from the observer coupled with the finite speed of light.

As our data are adequately explained by a quadratic function, we can provide only speculative limits on the properties of a sinusoid that might also provide a good fit to the timings. To do so, we fit a linear ephemeris to all timings obtained before JD 2456600 and found that the most recent data arrived $\sim$100\,s earlier than predicted by this ephemeris. As lower limits we adopted a period of triple the time interval covered by all of the timings, and an amplitude of 300\,s. This yields a mass function of $2 \times 10^{16}$\,kg, giving a minimum mass of the putative third component of {$\sim$}$2 \times 10^{-5}$\Msun\ (0.02\Mjup). An object of this mass would have easily evaded spectroscopic detection, and at a minimum orbital separation of 0.6\,au it would be much too close for detection via high-resolution imaging.

If present, this third object would imprint an orbital motion of amplitude $\sim$500\ms\ on the radial velocity (RV) of the star. \citet{Knutson+14apj} searched for changes in the systemic velocity of WASP-4\,A, finding an insignificant value of $\dot\gamma = -0.0099^{+0.0052}_{-0.0054}$\,m\,s$^{-1}$\,d$^{-1}$. The 3$\sigma$ upper limit on this is \reff{5.9\ms\,yr$^{-1}$, so would be} 36\ms\ over the 6-year time interval covered by the RVs and 69\ms\ over the 11.7-year time interval covered by the transit timings. The lowest possible RV change over a 6-year time interval is 67\ms, two times larger than allowed by the 3$\sigma$ upper limit on $\dot\gamma$, so the third-body interpretation is moderately inconsistent with the data.

Additional possibilities that could be invoked to bring the third-body hypothesis into better agreement with the data are orbital eccentricity (which must be sufficiently low to yield a roughly quadratic variation over the 11.7-year interval covered by the available timings), an inclined orbit (which will not help much because it would require a larger third body mass to obtain the same amplitude of the light-time effect), and a longer orbital period (which would also require a more massive third body). We conclude that the light-time effect is not a good explanation for the observed TTVs. Further analysis and RV measurements of the host star would help to strengthen this conclusion.




\subsection{Apsidal motion}

\citet{Patra+17aj} attempted to explain the quadratic orbital ephemeris of WASP-12 through apsidal precession, finding that it was consistent with the data but disfavoured relative to an orbital decay model. It is difficult to apply this model to WASP-4 because it requires measurements of the midpoint of occultations as well as transits over a significant time interval. Whilst some occultation timings are available \citep{Beerer+11apj,Caceres+11aa,Zhou+15mn} they lack the precision and time sampling necessary to provide useful constraints on apsidal motion in the system (see also B19). We recommend that new high-precision observations of occultations in the WASP-4 system are obtained.

Apsidal motion can only occur in eccentric systems. \citet{Triaud+10aa} found $e < 0.018$ for WASP-4 and \citet{Beerer+11apj} found $e\cos\omega = 0.00030 \pm 0.00086$: these do not constrain eccentricity sufficiently to rule out apsidal precession. One problem for this hypothesis, however, is the need for the system to have a non-zero eccentricity despite the presence of strong tidal effects in this short-period system containing a \reff{cool star with an extensive convective envelope.}


\section{Summary}

We have presented 22 times of mid-transit for the WASP-4 transiting planetary system that confirm the early arrival of transits noticed in TESS data (B19) and are consistent with a quadratic ephemeris. A constant-orbital-period model is rejected at the \refff{8.4}$\sigma$ significance level, making WASP-4\,b only the third hot Jupiter with TTVs detected to high significance.

We have explored possible explanations for the observed TTVs. Instrumental timing effects cannot be blamed because there is good agreement between different telescopes in the same observing seasons, and they are also not apparent in any of the tests we have performed. Biases due to spot crossing events in the light curves have a negligible effect on our results so can be ruled out as the source of the TTVs. The Applegate mechanism gives TTVs a factor of six smaller than observed. The light-time effect due to a third body struggles to match our results. Apsidal precession is plausible and deserves further investigation, in particular the observation of secondary eclipses so the orbital period found from transits can be compared to that found from occultations.

\reff{Orbital decay is a plausible but not favourable option for causing the observed TTV. The tidal quality factor needed, $Q_\star^\prime = 10^{4.58 \pm 0.03}$, is smaller than those predicted from theoretical arguments or measured from the population of known transiting planets. However, the envelope of possible values of $Q_\star^\prime$ is broad and close to the value found for WASP-4, so it is not currently possible to discount this hypothesis.}

\reff{A program of transiting timing measurements of a wide variety of planetary systems would help by providing observational constraints on orbital decay as a function of stellar type, planet mass, and orbital period. As $Q_\star^\prime$ is expected to depend on the system parameters, it also would be interesting to perform TTV analyses on other systems similar to WASP-4. We used the TEPCat\footnote{TEPCat (the Transiting Extrasolar Planet Catalogue) is available at {\tt http://www.astro.keele.ac.uk/jkt/tepcat/}} catalogue \citep{Me11mn} to identify HATS-2 and WASP-64 as the best candidates. These were found 5--6 years later than WASP-4 so their transit timings cover a smaller time interval, but this could be compensated for by obtaining a larger number of high-quality transit observations in the near future.}

We \reff{also} advocate obtaining new observations of the WASP-4 system, and will be keeping it on the target list of the MiNDSTEp transit monitoring program. New transit times would refine and extend the baseline of the quadratic ephemeris. Further timings of secondary eclipses would allow the apsidal motion hypothesis to be ruled in or out. RV measurements of the planet host star would give stricter limits on the change in systemic velocity of the system. WASP-4 and WASP-12 stand out as the only hot Jupiters for which unexplained TTVs have been detected: as this phenomenon is uncommon it is reasonable to propose unusual or unlikely scenarios for such systems.


\section*{Acknowledgements}

This paper uses data from the South African Astronomical Observatory (SAAO) under programmes Southworth-2016-05-1.0-m and Southworth-2019-01-40-inch-228, and from ESO under programme ID 088.C-0204 (P.I.\ Tregloan-Reed). The reduced light curves presented in this work will be made available at the CDS ({\tt http://cdsweb.u-strasbg.fr/}) and at {\tt http://www.astro.keele.ac.uk/jkt/}. We are grateful to Luke Bouma, Adrian Barker and an anonymous referee for very useful comments that helped to significantly improve our paper.
This project was partially funded by the MINEDUC-UA ESR project ANT 1795.
The following internet-based resources were used in research for this paper: the ESO Digitized Sky Survey; the NASA Astrophysics Data System; the SIMBAD database operated at CDS, Strasbourg, France; and the ar$\chi$iv scientific paper preprint service operated by Cornell University.


\bibliographystyle{mn_new}

\end{document}